\RequirePackage{lineno}
\documentclass{nature}
\usepackage{xcolor}
\usepackage{amsmath}
\usepackage{amssymb}
\usepackage{caption}
\usepackage{graphicx}
\usepackage{tabularx}
\usepackage{makecell}
\usepackage{physics}
\newcommand{\equa}[1]{Eq.~\eqref{#1}}

\renewcommand{\thefigure}{\textbf{Fig.~\arabic{figure}}}
\usepackage{float}
\makeatletter
\let\saved@includegraphics\includegraphics
\AtBeginDocument{\let\includegraphics\saved@includegraphics}
\renewenvironment*{figure}{\@float{figure}}{\end@float}
\makeatother
\usepackage{booktabs}
\usepackage{tabularx}
\usepackage{array}

\newcommand{\rom}[1]{\uppercase\expandafter{\romannumeral #1\relax}}

\usepackage{ulem}

\title{Observation of light-induced Haldane gap in graphene}

\author{Fei Wang$^{1,2,\ast}$, Xuanxi Cai$^{1,2,\ast}$,  Wanying Chen$^{1,2,\ast}$, Xiao Tang$^{1,2}$, Jinze Li$^{1,2}$, Benshu Fan$^{3}$, Tianshuang Sheng$^{1,2}$, Jinxi Lu$^{1,2}$, Jiansong Li$^{1,2}$, Hongyun Zhang$^{1,2,4}$, Pu Yu$^{1,2,5}$, Peizhe Tang$^{3,6}$, Wenhui Duan$^{1,2,5}$, Angel Rubio$^{3,7,8}$, Zhiyuan Sun$^{1,2,5}$ \& Shuyun Zhou$^{1,2,5,\dagger}$}

\usepackage{graphicx}
\makeatletter
\let\saved@includegraphics\includegraphics
\AtBeginDocument{\let\includegraphics\saved@includegraphics}
\renewenvironment*{figure}{\@float{figure}}{\end@float}
\makeatother
\begin{document}
\maketitle

\begin{figure}[H]
	\centering
	\includegraphics[
	width=0.75\linewidth,
	height=5.5cm,
	keepaspectratio
	]{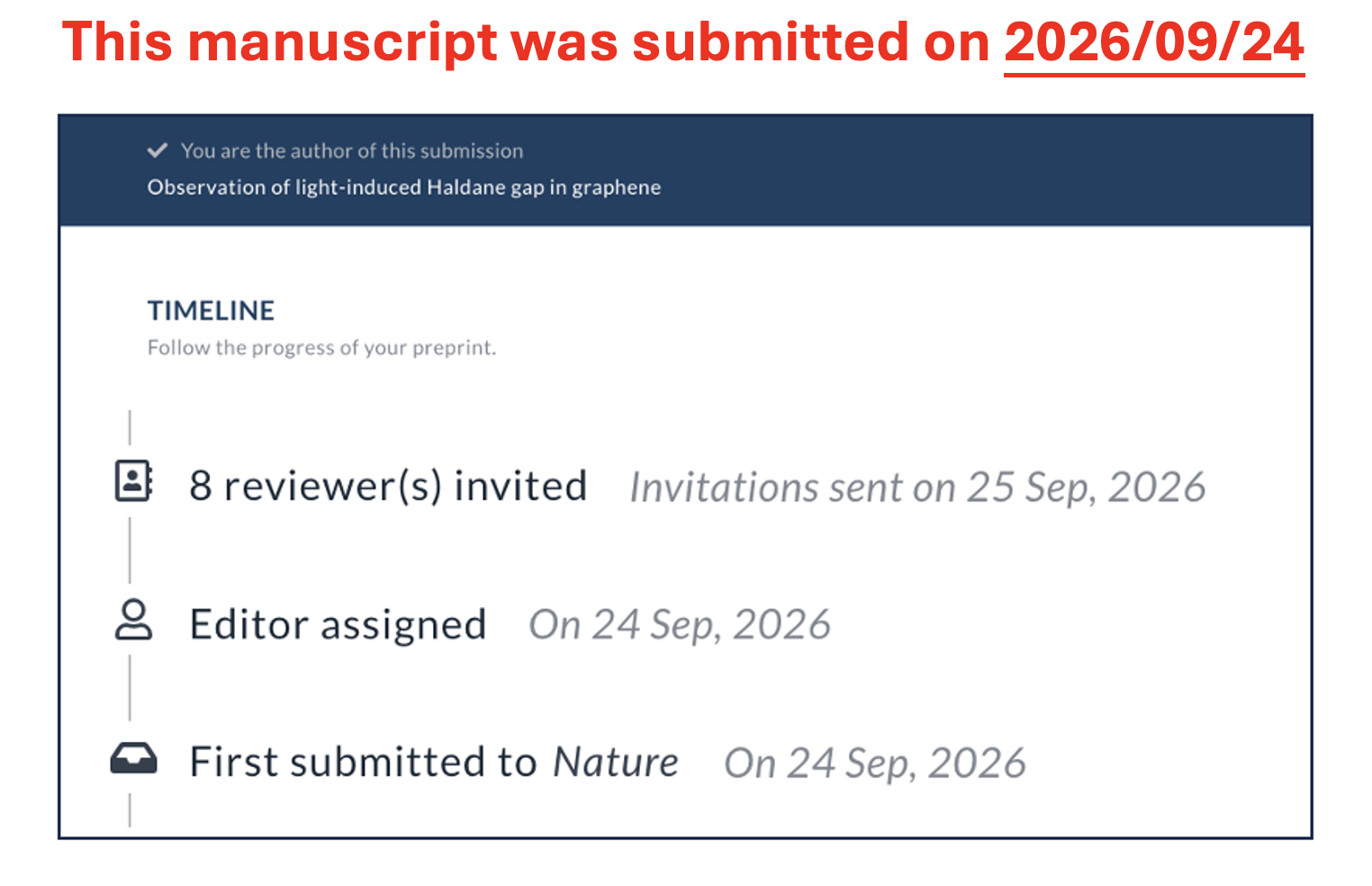}
\end{figure}

\begin{affiliations}

\item Department of Physics, Tsinghua University, Beijing 100084, People's Republic of China
\item State Key Laboratory of Low-Dimensional Quantum Physics, Tsinghua University, Beijing 100084, People's Republic of China
\item Max Planck Institute for the Structure and Dynamics of Matter, Luruper Chaussee 149, 22761 Hamburg, Germany.
\item Beijing Tsinghua Frontier Interdisciplinary Innovation Research Institute, Beijing, China
\item Frontier Science Center for Quantum Information, Beijing 100084, People's Republic of China
\item School of Materials Science and Engineering, Beihang University, Beijing 100191, People’s Republic of China
\item Initiative for Computational Catalysis and Center for Computational Quantum Physics, Simons Foundation Flatiron Institute, New York, New York 10010, USA
\item Nano-Bio Spectroscopy Group and ETSF, Universidad del País Vasco UPV/EHU- 20018 San Sebastián, Spain

* These authors contributed equally to this work\\
$\dagger$ Correspondence and request for materials should be sent to syzhou@mail.tsinghua.edu.cn

\end{affiliations}

\begin{abstract}

Symmetry and topology, encoded in electronic wavefunctions, play a central role in the electronic properties and emergent phenomena of quantum materials. While these properties are conventionally determined by the equilibrium Hamiltonian, strong optical fields offer a fundamentally different route by reshaping the Hamiltonian on ultrafast timescales, turning symmetry and topology into dynamically controllable degrees of freedom. 
Graphene provides a paradigmatic platform for realizing this concept, where circularly polarized light has long been predicted to dynamically break time-reversal symmetry (TRS) and open a topologically nontrivial gap at the Dirac point. 
Despite extensive experimental efforts, direct observation of this Dirac gap—the defining spectroscopic signature of the dynamical TRS-breaking induced topological phase—has remained missing. 
Here we directly observe this light-induced Dirac gap in monolayer graphene. Time- and angle-resolved photoemission spectroscopy reveals that the gap emerges simultaneously with the optical field and exclusively under circularly polarized driving, directly linking its opening to dynamical TRS breaking. Together with Floquet theory, we identify the gap with a light-induced Haldane mass and reveal the associated nontrivial Floquet Berry curvature and Chern topology. 
By demonstrating dynamical TRS breaking in a prototypical non-magnetic material, our results establish optical driving as a route to ultrafast engineering of symmetry and topology, opening opportunities for dynamically tunable quantum matter beyond equilibrium.

\end{abstract}

Symmetry and topology are among the central concepts of the electronic structures and emergent phenomena of quantum materials\cite{TKNN1982PRL,haldane1988model,kane2005prl}. Symmetries constrain the Hamiltonian and protect characteristic band degeneracies, 
while topology characterizes the global structure of electronic wavefunctions, giving rise to robust physical properties that are insensitive to microscopic details.
In equilibrium, these properties are largely governed by crystal structure, magnetic order, static or quasistatic external fields. Of particular importance for electronic topology is time-reversal symmetry (TRS): breaking TRS removes the constraint on the net Berry curvature flux and enables nontrivial Chern topology, as exemplified by quantum Hall\cite{vonKlitzing1980PRL}, quantum anomalous Hall\cite{XueScience2013,ZhangYBScience2020} and fractional quantum anomalous Hall\cite{XuXDFQAHENature2023,ShanJNature2023,LiTXPRX2023,JuLongNature2024} systems.

Coherent light fields provide a new route to dynamically control TRS by engineering the electronic Hamiltonian on ultrafast timescales through strong light-matter coupling\cite{oka2009PRB}. In the long-wavelength dipole approximation, the vector potential $\mathbf{A}(t)$ of the optical field enters the Bloch Hamiltonian as $H_0(\mathbf{k})\rightarrow H_0[\mathbf{k}+e\mathbf{A}(t)/\hbar]$, where $e$ is the elementary charge and $\hbar$ is the reduced Planck constant. By tuning the light polarization, frequency and amplitude, time-periodic driving provides a dynamical pathway to reshape the electronic band structures\cite{GedikSci2013,HuberNature2023,zhou2023pseudospin} with light-induced emergent phenomena\cite{HsiehNM2017,oka2019floquet,Rudner2020Floquet,SentefRMP2021,ZhouNRP2022}. Circularly polarized light is particularly intriguing because its handedness offers a route to dynamically break TRS even in a non-magnetic material (Fig.~1a). Under periodic driving, this symmetry breaking is encoded in an effective Hamiltonian\cite{oka2009PRB} and has been predicted to endow the resulting Floquet states with nontrivial topology absent at equilibrium\cite{Demler2011prb,lindner2011NP}. 

\renewcommand{\thefigure}{\textbf{Fig.~\arabic{figure}}}
\setcounter{figure}{0}
\begin{figure*}
	\centering
	\includegraphics[width=16.8cm]{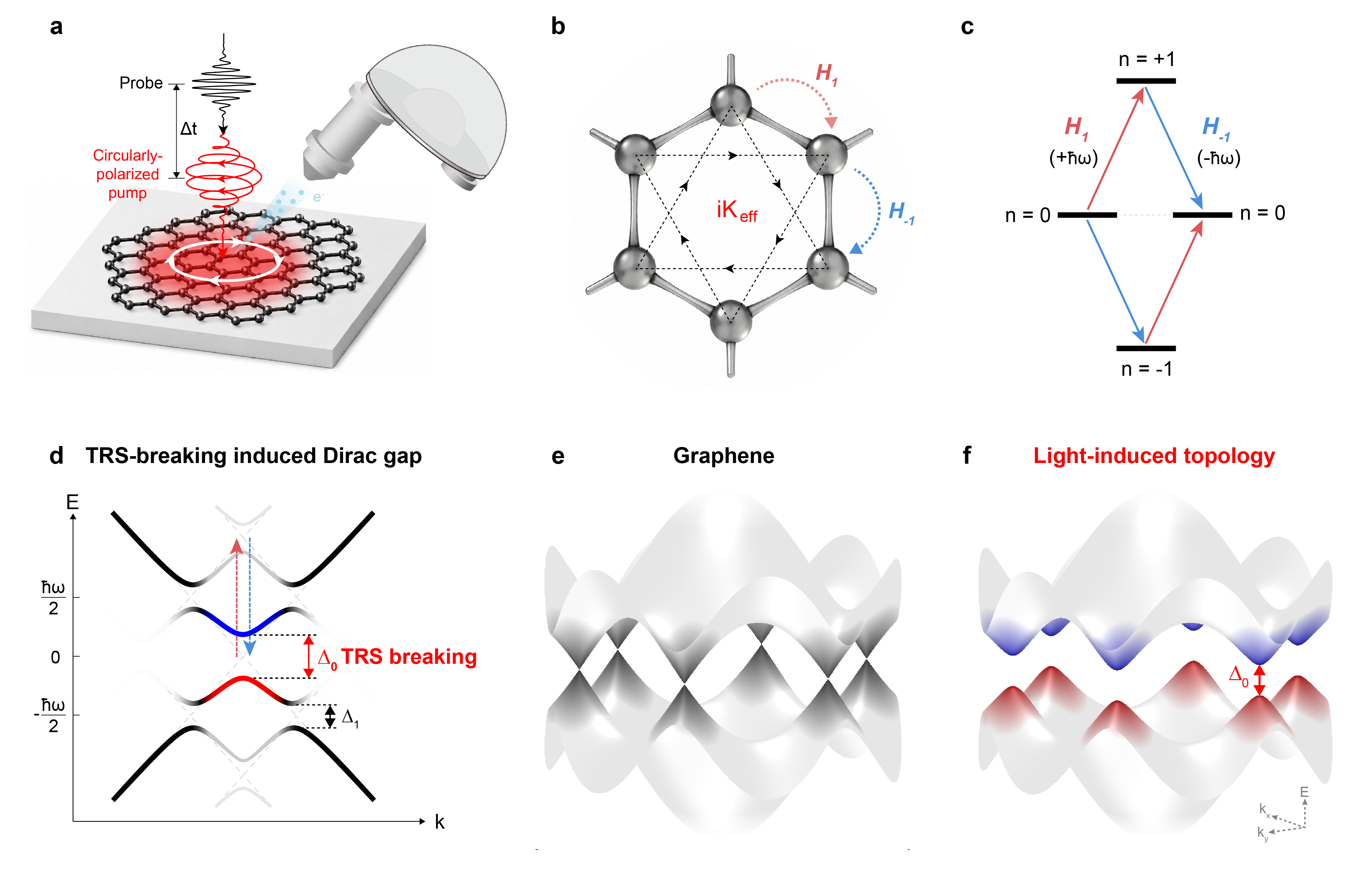}
	\caption*{{\bf Fig.~1 ${\mid}$  Schematic illustration of Floquet topological phases in irradiated graphene upon circularly polarized driving.}
	 {\bf a}, Monolayer graphene under circularly polarized driving.
	 {\bf b}, The effective complex next-nearest-neighbor hopping \(iK_{\mathrm{eff}}\) induced through virtual photon absorption-emission process.
	 {\bf c}, The Hamiltonian modification arises from a two-photon absorption-emission hopping process.
	 {\bf d}, Schematic electronic structure of irradiated graphene under circularly polarized driving. 
	 {\bf e,f}, Schematic electronic structure of equilibrium graphene ({\bf e}) and irradiated graphene under circularly polarized driving in the high frequency limit (\textbf{f}).
   }
\label{Fig1}
\end{figure*}

Graphene provides a paradigmatic system to test this possibility, whose massless Dirac fermions are protected by inversion symmetry and TRS\cite{NetoRMP09}. The Haldane model\cite{haldane1988model} demonstrated theoretically that TRS breaking can generate a topological gap and a Chern insulating state without a net magnetic flux. 
As a dynamical alternative to static TRS breaking, circularly polarized light was proposed to realize an analogous mechanism in graphene\cite{oka2009PRB}. Second-order virtual photon absorption-emission processes, encoded by the $n=\pm1$ Fourier components $H_{\pm1}$ of the time-periodic Hamiltonian, generate an effective TRS-breaking mass term in the low-energy Floquet Hamiltonian\cite{Demler2011prb,Devereaux2015NC},
$\frac{[H_{-1},H_{+1}]}{\hbar\omega}=m_{\mathrm F}\sigma_z$,
where $m_{\mathrm F}$ denotes the light-induced Floquet mass and $\sigma_z$ is the Pauli matrix acting on the graphene sublattice pseudospin (Fig.~1b,c).
It thus opens a Dirac gap and generates nonzero Floquet Berry curvature (Fig.~1d-f).
 This direct correspondence between symmetry breaking, Dirac-mass generation and topology makes graphene a canonical system for realizing light-induced topological state.

Experimental progress toward Floquet topological engineering in graphene has advanced from the detection of dressed Floquet states\cite{GierzNanoLett2021,lee2022nat,stefan2025graphene,gedik2025graphene,DaniNPNV2025}, to light-induced reconstruction of the electronic structure via the linearly polarized driving\cite{zhou2026graphene,zhou2026graphite,SentefNM2026}. Crucially, the recent observation of avoided-crossing gaps at Floquet resonances 
established that experimentally accessible optical fields can directly reshape the electronic structure of graphene\cite{zhou2026graphene}. A central goal of Floquet topological engineering, however, is to move beyond electronic structure reconstruction and dynamically control the topology of electronic wavefunctions.
In graphene, it is proposed that that circularly polarized light can achieve this by dynamically breaking TRS and opening a Haldane gap at the Dirac point\cite{oka2009PRB,Demler2011prb,Devereaux2015NC,RubioSciAdv2025}.
Experimentally, light-induced anomalous Hall responses\cite{Cavalleri2020np} and dichroic photocurrents\cite{Hommelhoff2026np} have provided important signatures consistent with this scenario. 
However, these measurements probe the driven response rather than the topological electronic structure itself. Because such signals are strongly shaped by nonequilibrium carrier populations, scattering and dissipative dynamics\cite{oka2009PRB,Demler2011prb, SatoNJP2019,RubioPRB2019,MatheyPRR2020}, they cannot by themselves establish a TRS-breaking topological gap.
The decisive missing evidence is a direct, energy- and momentum-resolved observation of the Dirac gap---the spectroscopic manifestation of light-induced Haldane mass. Such an observation would directly establish the emergence of non-trivial topology under circularly polarized driving, a long-standing goal of Floquet engineering in graphene.

Here by accessing the strong light--matter coupling regime, we directly observe this light-induced Dirac gap in monolayer graphene using time- and angle-resolved photoemission spectroscopy (TrARPES). The gap is observed exclusively under circularly polarized driving and emerges synchronously with the optical field, supporting its dynamical TRS-breaking origin, while theoretical calculations reveal the associated Floquet Berry curvature and Chern topology. Together, these results establish that an optical field can dynamically break TRS, thereby creating Floquet Chern state in a non-magnetic material.

\section*{Signature of Floquet-induced Dirac gap from the temporal evolution}

\begin{figure*}
	\centering
	\includegraphics[width=16.8cm]{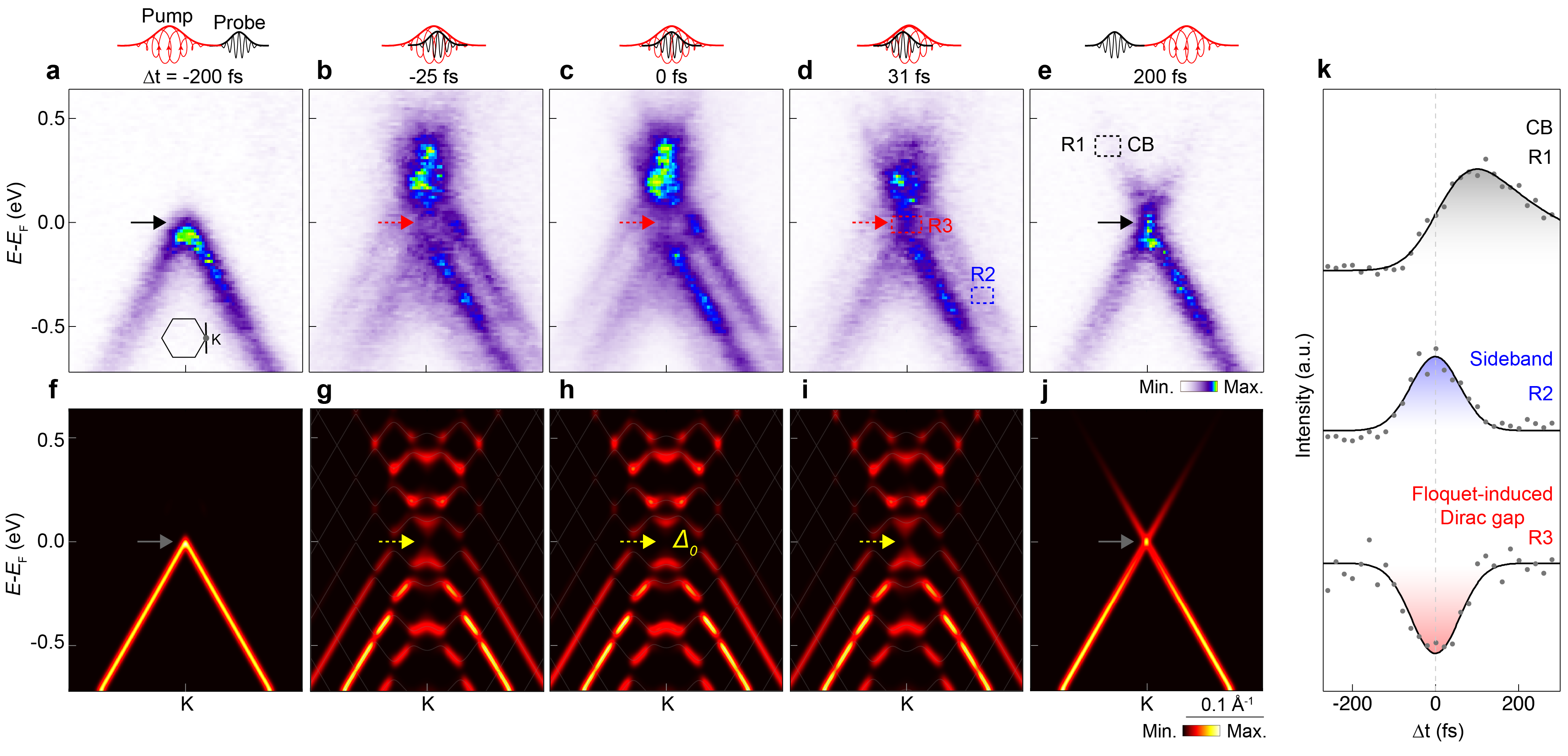}
	\caption*{{\bf Fig.~2 ${\mid}$  TrARPES signatures of light-induced electronic structure reconstruction in monolayer graphene.}
	{\bf a-e}, Dispersion images measured at different delay time $\Delta t$ under circularly polarized driving with $\hbar\omega=310$ meV at a pump fluence of 9.0 mJ/cm$^{2}$. The measurement direction is indicated by the black line in the inset.
	{\bf f-j}, Simulated TrARPES spectra at different delay times. 
	{\bf k}, Integrated intensity of the photo-excited CB (1), sideband (2) and the Dirac point (3) as a function of delay time over regions marked in (d) and (e). 
}\label{Fig2}
\end{figure*}

Unambiguous identification of the light-induced Dirac gap $\Delta_0$ requires an intrinsically gapless monolayer graphene. In epitaxial graphene, the substrate-induced equilibrium gap\cite{syzhou2007NM,SonYWPRL08} and mid-gap states can obscure the genuine light-induced gap. We therefore use a large, highly uniform exfoliated monolayer graphene flake, enabling high-quality TrARPES data to reveal the sharp, gapless Dirac cone
. Resolving the topological Dirac gap further requires a qualitatively different driving regime from that used to observe first-order Floquet hybridization gaps\cite{zhou2026graphene,zhou2026graphite,zhou2023pseudospin}. Because the TRS-breaking gap arises from a higher-order Floquet effect\cite{oka2009PRB}, here we combine a low pump photon energy of $\hbar\omega=310$ meV with typical fluence of 9 mJ/cm$^2$ (in-plane field strength  $1.0 \times10^{8}$ V/m) to enhance this interaction and bring the light-induced Dirac gap into an experimentally resolvable regime.

Figure~2 tracks the transient electronic structure across the pump--probe delay under circularly polarized driving. Before the pump pulse ($\Delta t=-200$ fs), graphene exhibits its equilibrium gapless Dirac dispersion with the Dirac point located at the Fermi level (Fig.~2a). When the pump overlaps with the probe (Fig.~2b-d), the electronic structure undergoes pronounced reconstruction, characterized by the Floquet sidebands separated by the pump photon energy and, more importantly, a strong suppression of spectral weight near the K point, providing the first indication of a transient Dirac gap. At 200 fs after the pump pulse, both features disappear and the Dirac cone dispersion is restored, while the upper Dirac cone remains transiently populated (Fig.~2e). Calculated spectra from Floquet theory reproduce these essential features, including the Floquet sidebands and the suppression of spectral weight near the Dirac point (Fig.~2f-j), supporting that they both arise from coherent coupling to the light field.

Because photo-excited carriers and coherent light-field dressing coexist during excitation, we distinguish their contributions through their distinct temporal dynamics. Figure~2k shows the evolution of the integrated intensities of the photo-excited conduction band (CB), the Floquet sideband and the spectral weight near the Dirac point. Whereas the photo-excited CB population persists over several hundred femtoseconds before fully relaxing, both the Floquet sideband and the suppression of the spectral weight at the Dirac point are confined only to the pump--probe temporal-overlap region, distinguishing them from long-lived population effects. Their similar temporal evolution further identifies the intensity suppression near the Dirac point as a coherent Floquet response associated with the light-induced Dirac gap.

\section*{Extraction of light-induced Dirac gap under circularly polarized driving}

\begin{figure*}
	\centering
	\includegraphics[width=16.8cm]{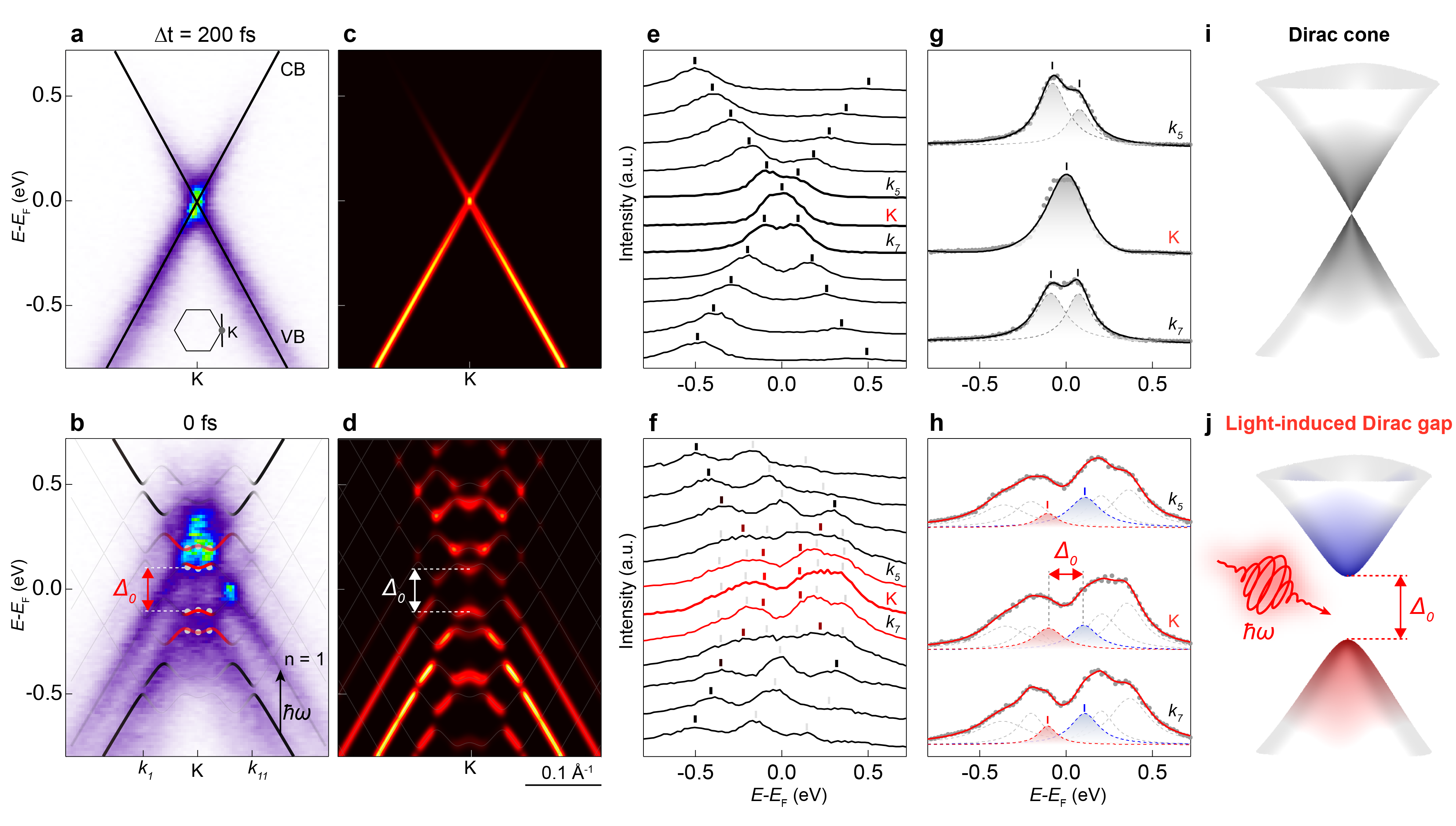}
	\caption*{{\bf Fig.~3 ${\mid}$  Experimental observation of Floquet-induced Dirac gap in monolayer graphene.}
	{\bf a,b}, Dispersion images measured at $\Delta t=200$ fs ({\bf a}) and $\Delta t=0$ fs ({\bf b}) under circularly polarized driving with $\hbar\omega=310$ meV at a pump fluence of 9.0 mJ/cm$^{2}$.
	{\bf c,d}, Simulated TrARPES spectra at $\Delta t=200$ fs ({\bf c}) and $\Delta t=0$ fs ({\bf d}). 
	{\bf e,f}, EDCs extracted from ({\bf a}) and ({\bf b}) at the momentum positions marked in ({\bf b}). The tick marks indicate the peak positions.
	{\bf g,h}, EDCs at representative momentum positions ($k_5$, K and $k_7$) with fitting curves appended.
	{\bf i,j}, Schematics of the gapless Dirac cone ({\bf i}) and the Floquet-induced Dirac gap upon circularly polarized driving ({\bf j}).
}\label{Fig3}
\end{figure*}

Figure~3 directly resolves the light-induced Dirac gap by comparing the electronic structure at $\Delta t=200$~fs and $\Delta t=0$. At $\Delta t=200$~fs, TrARPES resolves the intrinsic gapless Dirac cone (Fig.~3a). At temporal overlap under circularly polarized driving (Fig.~3b), a pronounced suppression of spectral weight near the K point is observed, accompanied by a clear separation of the upper and lower Dirac branches, providing direct spectroscopic evidence for gap opening at the Dirac point (Fig.~3b). Floquet calculations reproduce both the gap and the accompanying sidebands (Fig.~3c,d), supporting their light-induced origin.

The gap magnitude is further extracted from energy distribution curve (EDC) analysis (Fig.~3e-h). Whereas the EDC at $\Delta t=200$~fs shows a single peak at the Dirac point, a clear splitting is observed at $\Delta t=0$ (see the schematic comparison in Fig.~3i,j). Tracking the momentum-dependent peak positions allows us to extract the gapped dispersion (tick marks in Fig.~3f), where the separation of the upper and lower branches at the K point defines $\Delta_0$. Fitting these peak positions gives the gap value $\Delta_0=203\pm30$ meV (Fig.~3h). 

\section*{Polarization-selective dynamical TRS breaking}

\begin{figure*}
	\centering
	\includegraphics[width=16.8cm]{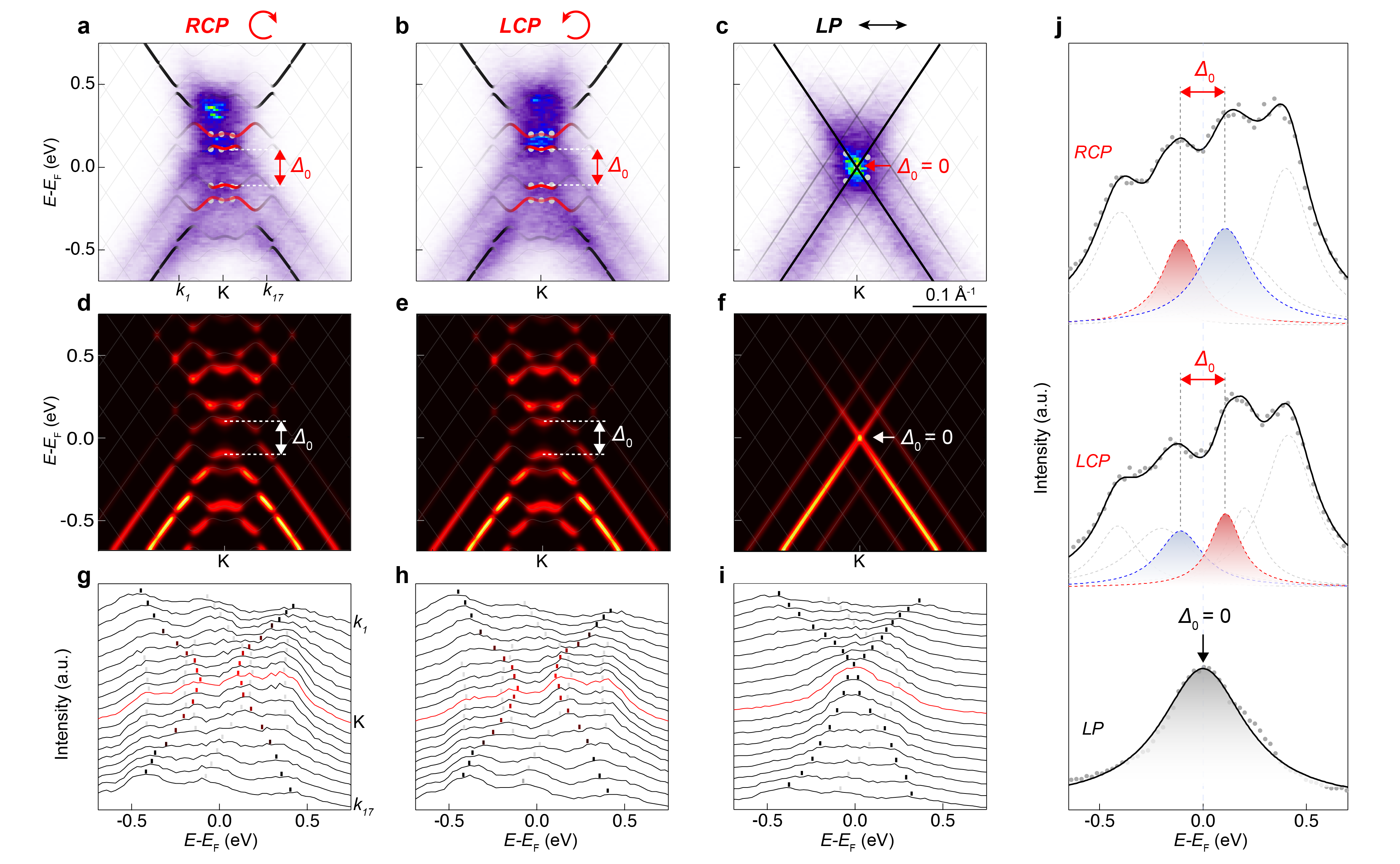}
	\caption*{{\bf Fig.~4 ${\mid}$ Floquet-induced TRS-breaking gap observed exclusively under circularly polarized driving.}
   {\bf a-c}, TrARPES dispersions measured under right- and left- circularly polarized driving ({\bf a,b}) and linearly polarized driving ({\bf c}). The pump photon energy is 310 meV and the pump fluence is 10 mJ/cm$^2$. 
   {\bf d-f}, Corresponding simulations for different polarized driving.
	{\bf g-i}, EDCs extracted from ({\bf a-c}) at the momentum positions marked in ({\bf a}). The tick marks indicate the peak positions.
	{\bf j}, EDCs at the K point with fitting curves appended.
	}\label{Fig4}
\end{figure*}

The dynamical TRS-breaking origin of the Dirac gap is established by its polarization dependence. Figure 4a-c compares the dispersion images measured at $\Delta t = 0$ under right- and left-circularly polarized (RCP and LCP) driving and linearly polarized (LP) driving with the light field parallel to the electron momentum. A pronounced Dirac gap opens under both RCP and LCP driving but is absent under LP excitation. Instead, LP driving results in only anisotropic avoided-crossings at Floquet resonances, consistent with previous observation in monolayer graphene\cite{zhou2026graphene}and graphite\cite{zhou2026graphite}. Floquet simulations in Fig.~4d-f reproduce this polarization selectivity, with a Dirac gap emerging exclusively under circular driving. Together, experimental and theoretical results identify the gap as a direct signature of dynamical TRS breaking. 

The polarization dependence is further quantified by EDC analysis in Fig.~4g-j. Fitting the EDCs at the K point yields nearly identical gap magnitudes for RCP and LCP drivings (Fig.~4j), consistent with the Floquet TRS-breaking mechanism\cite{oka2009PRB}, where helicity reversal changes the sign of the induced Dirac mass while preserving its magnitude. Together, the exclusive emergence of the Dirac gap under circularly polarized driving and its helicity-independent magnitude distinguish the second-order TRS-breaking mass from first-order Floquet hybridization, establishing the observed gap as a direct spectroscopic signature of dynamical TRS breaking.

\section*{Light-induced Haldane mass and Floquet Berry curvature}

\begin{figure}\label{Fig5}
	\centering
	\includegraphics[width=16.8cm]{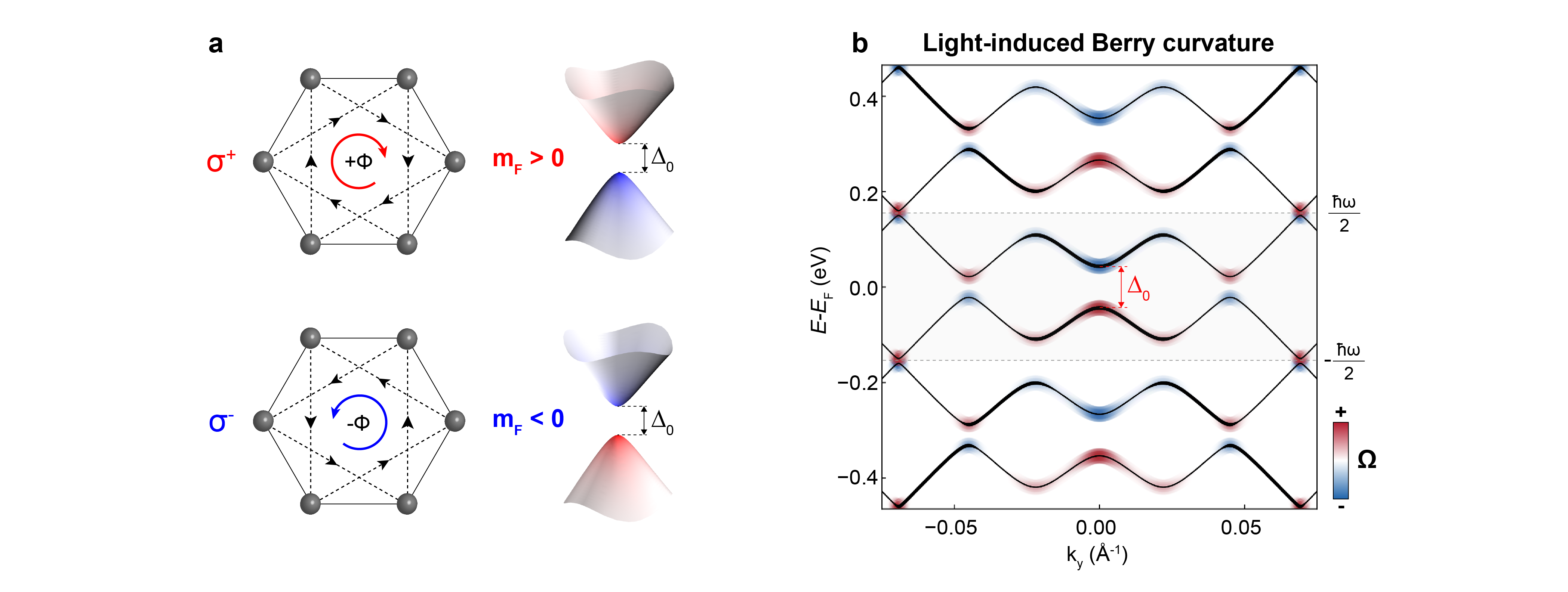}
	\caption*{{\bf Fig.~5 ${\mid}$ Light-induced Haldane mass and Berry curvature under circularly polarized driving.}
      {\bf a}, Schematic of RCP pump generating an effective chiral next-nearest-neighbor hopping, producing a Haldane mass, whose sign reverses for LCP pump.
      {\bf b}, Theoretical Floquet electronic bands\cite{oka2009PRB} with the Floquet Berry curvature encoded by the color scale.
}\end{figure}

The observation of the Dirac gap exclusively under circularly polarized driving provides direct evidence for a TRS-breaking Haldane mass underlying the Floquet topological bands in graphene\cite{haldane1988model,oka2009PRB,Demler2011prb}. 
In  the high frequency expansion for the effective Hamiltonian, circularly polarized light generates Dirac masses of opposite sign at the K and K$^\prime$ valleys, with both signs reversed upon reversing the light helicity (Fig.~5a). 
In real space, this mass arises from effective complex next-nearest-neighbor hoppings (schematically illustrated in Fig.~1b) generated through second-order virtual photon processes, directly analogous to the Haldane model\cite{haldane1988model,Demler2011prb}. 
This low-energy description applies near the Dirac points for electronic energies well below the pump photon energy $\hbar\omega$. 
Combined with the opposite Dirac chiralities between the K and K$^\prime$ valleys, the valley-opposite masses generate Floquet Berry curvatures of the same sign, providing the characteristic topological signature\cite{Demler2011prb} analogous to the Haldane model (see Fig.~5 and Extended Data Fig.~1). Our measurements and calculations therefore establish a light-induced Haldane mass near the Dirac points and the associated low-energy Floquet Chern topology.

\section*{Discussion and perspectives}

Our work establishes dynamical TRS breaking as a route to creating nontrivial electronic topology in a non-magnetic material. Previous realizations of light-induced TRS breaking have largely been limited to synthetic platforms\cite{Esslinger2014coldatom,Szameit2013photonic} or to solids with pre-existing nontrivial topology or magnetic order\cite{GedikSci2013,MahmoodNP2025,monneySnTe2026np}. 
Here, by contrast, optical driving generates a topological phase from an initially gapless, non-magnetic electronic system, driving graphene into a Chern state. 
More broadly, these results show that a fundamental symmetry conventionally determined by the equilibrium Hamiltonian can become a transient, optically tunable degree of freedom.

Dynamical TRS breaking provides a general route to optically transforming topological states, for example,  driving Dirac or nodal-line semimetals into Weyl phases\cite{LeePA2016PRB,WangZ2016PRL,TaguchiPRB2016,Hubener2017NC,Yuta2023prl}. More fundamentally, it is not simply an optical analogue of a magnetic field: unlike static magnetic fields, it can break TRS without requiring magnetic order or spin polarization, providing access to symmetry-broken states beyond conventional magnetic approaches. This distinct route of symmetry control may open access to chiral many-body states that are difficult to
realize with static magnetic fields, for example, chiral spin liquids\cite{Claassen2017NC,Kitamura2017PRB}. Moreover, combining light-induced Chern bands with strong electronic correlations may enable their fractionalization into Floquet fractional Chern insulating states\cite{Neupert2014FFCI}, while related chiral interactions have been proposed to generate Floquet topological superconductivity\cite{Kitamura2022CP}. Dynamical TRS breaking thus opens a new pathway for creating and controlling chiral quantum matter beyond equilibrium.

\begin{methods}




\section*{Experimental setup}

TrARPES measurements were performed using a home-built system at Tsinghua University. The laser source was a regenerative amplifier delivering 35 fs pulses centered at 800 nm (1.55 eV), with a pulse energy of 1.3 mJ and a repetition rate of 10 kHz. Probe pulses at 21.7 eV were generated by high-order harmonic generation (HHG) in argon driven by frequency-doubled 800 nm light and isolated using aluminium and tin filters\cite{syzhou2024hhg}. The resulting probe pulses had a duration of 66 fs, providing sufficient temporal resolution to capture the Floquet-induced electronic reconstruction before electron-phonon and other relaxation processes become prominent\cite{Cavalleri2013NM}. The probe beam was focused onto the sample using a gold-coated toroidal mirror, producing a tightly-focused spot size of $57\,\mu{\rm m}\times90\,\mu{\rm m}$ and enabling measurements of mechanically exfoliated graphene flakes.

Mid-infrared pump pulses with photon energy of 310 meV (wavelength of 4 \(\mu\)m) and pulse duration of 121 fs were generated by difference-frequency generation between the signal and idler outputs of an optical parametric amplifier\cite{zhou2023pseudospin}. Circularly polarized pump pulses were generated by passing through a quarter-wave plate designed for a center wavelength of 4 \(\mu\)m. The pump beam was incident nearly normal to the sample surface, at an angle of $0^\circ \pm 1^\circ$, ensuring a predominantly in-plane electric field. The degree of circular polarization was measured to be $95\% \pm 2\%$. The combination of high polarization purity and normal incidence geometry produces a highly circular, predominantly in-plane driving field, providing the conditions required for dynamical TRS breaking. In the TrARPES experiments, a typical pump fluence of 9.0 mJ/cm$^2$ was applied to MLG, which corresponds to an electric field of $1.0 \pm 0.2\times10^8$ V/m at the sample after accounting for Fresnel reflection, following the calculations in previous work\cite{zhou2026graphene}. 

\section*{Extraction of the Floquet-induced TRS-breaking gaps}

To quantify the Floquet-induced Dirac gap at the K point, each EDC was obtained by integrating the spectral intensity over a momentum window of 0.014~\AA$^{-1}$. Each EDC was fitted with six Lorentzian peaks to capture the two split Dirac-point peaks and their accompanying Floquet sideband peaks. The gap size, $\Delta_0$, was defined as the energy separation between the two split peaks at the Dirac point. The uncertainty of the extracted gap was estimated as $\delta=\sqrt{\delta_{\rm fit}^2+\delta_{\rm sys}^2}$, where $\delta_{\rm fit}$ is the statistical fitting uncertainty and $\delta_{\rm sys}$ is the systematic uncertainty evaluated by varying the momentum-integration window.

\section*{Calculation of the Floquet electronic structure}

The Floquet electronic structures and time-resolved photoemission spectra were calculated using a tight-binding model of monolayer graphene. 
The equilibrium nearest-neighbour spinless Hamiltonian in the $A/B$ sublattice basis is
\begin{equation}
H_0(\mathbf p)
=
-t_0 \begin{pmatrix}
0&g(\mathbf p)\\
g^*(\mathbf p)&0
\end{pmatrix},
\quad
g(\mathbf p)
=
\sum_{j=1}^{3}
e^{i\mathbf p\cdot\delta_j}
\,,
\end{equation}
where 
$\delta_1=a_{\text{C}}(0,1)$, $\delta_2=a_{\text{C}}(\sqrt{3}/2,-1/2)$ and $\delta_3=a_{\text{C}}(-\sqrt{3}/2,-1/2)$ are the three vectors connecting nearest-neighbor carbon atoms, $a_{\text{C}}=1.42~\text{\AA}$ is the C--C bond length, and the hopping amplitude is set to $t_0=3.09~\mathrm{eV}$ corresponding to the Dirac  velocity $v_{\text{F}} \simeq 1\times10^6~\mathrm{m/s}$ at the K point.
The pump field is introduced through minimal coupling,
$H(\vb p, t)=H_0(\vb p+e\vb A(t)/\hbar)$, where 
$\vb A(t)=A_0(\cos\omega t,\eta\sin\omega t)$ is the oscillating vector potential for circularly polarized light with the optical helicity $\eta=\pm1$.
The `Hamiltonian' in Floquet space is 
$
(H_{\text{F}})_{mn}
=
H_{m-n}
+
m\hbar\omega\,\delta_{mn}\mathbb I
$
where
$
H_\ell
=
\frac{1}{T}
\int_0^T dt\,
e^{-i\ell\omega t}H(t)
\,
$, $m,n$ label the Floquet sectors and $T=2\pi/\omega$. 
The  electronic Floquet bands in Figs.~2-5 are  obtained by diagonalizing $(H_{\text{F}})_{mn}$.
Convergence in the truncation $m,n=-N,\ldots,N$ is achieved with $N=8$. 
In the next section, we discuss analytical understanding of the Floquet electronic structure.

\section*{Floquet Haldane mass and Berry curvature of the low energy electrons}

The Haldane mass and Berry curvature are most conveniently understood from the standard high frequency expansion for the effective Floquet Hamiltonian\cite{Demler2011prb}, which works for electrons close to the K points and at energies much lower than the pump frequency $\omega$.
Close to the K points, the equilibrium Hamiltonian for the low energy Dirac electrons is
$
H_{0,\tau}(\mathbf k)
=
\hbar v_{\text{F}}
\left(
\tau k_x\sigma_x+k_y\sigma_y
\right),
$
where the Pauli matrices $\sigma_i$ act in the sublattice space and
 $\tau=+1$ ($-1$) corresponds to K (K$^\prime$) valley. 
Under circularly polarized pump light with helicity $\eta=\pm1$, to the lowest non-trivial order in $1/\omega$, virtual photon absorption and emission give the effective Floquet Hamiltonian\cite{Demler2011prb}
\begin{equation}\label{eqn:H_eff}
H_{\mathrm{eff},\tau}(\mathbf k)
\approx 
H_{0,\tau}
+
\frac{[H_{-1},H_{+1}]}{\hbar\omega}
=
\hbar v_{\text{F}}
\left(
\tau k_x\sigma_x+k_y\sigma_y
\right)
+
\tau\eta
m_{\text{F}}
\sigma_z
\end{equation}
where $m_{\text{F}}=\beta^2\hbar\omega$ and 
$\beta=e v_F A_0/(\hbar\omega)=e v_F E_0/(\hbar\omega^2)$. 
The  Dirac gap is $\Delta_0=2|m_{\text{F}}|$.
From \equa{eqn:H_eff}, the Haldane  mass $m_\tau=\tau\eta m_{\text{F}}$ has opposite signs at K and K$^\prime$, while reversing the optical helicity reverses the signs of both. 
The Berry curvature of the conduction/valence band of $H_{\mathrm{eff},\tau}$  takes the analytical form
$
\Omega_{s,\tau}(\mathbf k)
	=
	-s \eta \,
	\frac{
		m_{\text{F}}\hbar^2v_{\text{F}}^2
	}{
		2\left(
		\hbar^2 v_{\text{F}}^2k^2+m_{\text{F}}^2
		\right)^{3/2}
	},
$
where $s=+1$ ($-1$) denotes the conduction (valence) band. 
Between the two valleys, the opposite Dirac chiralities and opposite masses  compensate, resulting in Berry curvatures of the same sign. 
It concentrates within the region of the  characteristic momentum scale $k_{\text{m}}=m_{\text{F}}/(\hbar v_{\text{F}})$.
One may define the ``partial Chern number'' 
$
C_\Lambda
=
\frac{1}{2\pi}
	\int_{|\mathbf k|<\Lambda}
	d^2k\,\Omega(\mathbf k)
$
by integrating the time-averaged Berry curvature over a circular momentum region with the cutoff $\Lambda$ around the K point.
For the valence band, one has
$
	C_\Lambda
	=
	\eta
	\left(
	1-
	\frac{1}{
		\sqrt{1+(\Lambda/k_{\text{m}})^2}
	}
	\right)
	\xrightarrow{\Lambda\gg k_{\text{m}}}
	\eta
$
where we have summed two valleys.
We note again that this model works only for low electrons close to the K points, while the exact topological characterization of the Floquet band is discussed in the next section.

\section*{Exact Floquet Berry curvature and Floquet topology}

The exact topological classification of the Floquet bands requires a shift from the equilibrium framework.
Strictly speaking, 
the Floquet Berry curvature of a Floquet band $\alpha$ with the Floquet eigenstates \(|\phi_{\alpha,\mathbf{k}}(t)\rangle\) is
$
	\Omega_\alpha(\mathbf{k},t)=
	-i\nabla_{\mathbf{k}}\times
	\langle\phi_{\alpha,\mathbf{k}}(t)|
	\nabla_{\mathbf{k}}|\phi_{\alpha,\mathbf{k}}(t)\rangle
	$, 
which can vary with time within each driving cycle\cite{lindner2011NP,Rudner2013}.
However, its integral over the full Brillouin zone, the Floquet Chern number $C_\alpha=\frac{1}{2\pi}\int_{\mathrm{BZ}}d^2\mathbf{k} \Omega_\alpha(\mathbf{k},t)$, is a topological invariant that is time-independent because smooth unitary evolution preserves the band topology\cite{lindner2011NP,Rudner2013,Rudner2020Floquet,SchulerPRX2020,Asgari2026}. 
Writing the Floquet eigen state wave function as
$\phi_{\alpha,\mathbf{k}}(t)=e^{-i\varepsilon_\alpha t}\sum_n u_{\alpha n ,\mathbf{k}} e^{-in \omega t}
$
where $n$ is the harmonic index, the time-averaged Berry curvature is calculated as
$
\Omega_\alpha(\mathbf k)
=
-2\,\mathrm{Im}
\sum_n
\left\langle
\partial_{k_x}u_{\alpha n}
\middle|
\partial_{k_y}u_{\alpha n}
\right\rangle
$. The corresponding partial Chern number $C_\Lambda$ develops a plateau approaching $+1$ ($-1$) for RCP (LCP) for a substantial range of $\Lambda$ (see Extended Data Fig.~1), indicating the Chern topology associated with the light-induced Haldane mass. 

We now discuss the implications of such Floquet Chern topology.
An integer Chern number does not imply a quantized Hall conductance as the Floquet bands are generally only partially occupied\cite{oka2009PRB,Demler2011prb,SatoNJP2019,RubioPRB2019,MatheyPRR2020}.
Its topological significance lies instead in the chirality of Floquet edge states. The net chirality of edge states crossing the quasienergy \(\varepsilon\) within a quasienergy gap is characterized by another topological invariant, the winding number \(W_\varepsilon\)\cite{Rudner2013} that is related to the Chern number. 
Note that these topological invariants are rigorously defined only globally over the whole Brillouin zone and are therefore sensitive to the weak higher-order resonances\cite{UsaiPRA2015}.
The experimentally relevant low-energy topology, however, remains clear: the observed Dirac gap is identified with a dynamically generated Haldane mass near the Dirac point, together with the associated Berry-curvature structure and low-energy chiral edge states\cite{Demler2011prb,Rudner2020Floquet}.

\section*{Simulation of the time-resolved photoemission signal}
In this section, we discuss  the numerical calculation of the TrARPES intensity:
\begin{equation}
	I_{\mathbf{k}}(\omega,\Delta t)
	=
	\sum_{ab}
	M_{a,\mathbf{k}}M_{b,\mathbf{k}}^*
	\int dt\,dt'\,
	s(t)s(t')
	e^{i\omega(t-t')}
	\mathcal V_{\mathbf p}(t,t')
	\left[
	-iG^<_{ab,\mathbf{k}}(t,t')
	\right].
\end{equation}
Here $a,b=A,B$ are the sublattice indices, 
$M_{a/b,\mathbf{k}}$ is the photoemission matrix element between the free space electronic state $\mathbf{p}$ and the Bloch state at crystal momentum $\mathbf{k}$ (equal to the in-plane momentum $\mathbf{p}_\parallel$ of the free space state in our experiment) formed from an $a$- or $b$-sublattice orbital, and $s(t)$ denotes the probe-pulse envelope which has a delay $\Delta t$  relative to the pump.
The lesser Green function is
\begin{equation}
	G^<_{ab,\mathbf{k}}(t,t')
	=
	iU_{\mathbf{k}}(t,t_0)\rho_{0,ab,\mathbf{k}} 
	U_{\mathbf{k}}^\dagger(t',t_0),
	\quad	
	U_{\mathbf{k}}(t,t')
	=
	\mathcal{T}\exp[-(i/\hbar)\int_{t'}^{t}d\tau\,H(\mathbf{k},\tau)]
\end{equation}
where $\rho_{0,ab,\mathbf{k}} =\langle c_{a,\mathbf{k}}^\dagger c_{b,\mathbf{k}} \rangle_0$
is the single particle density matrix determined by the equilibrium Fermi--Dirac distribution before the pump, 
$
H(\mathbf{k},\tau)
$ 
is the time dependent single particle Hamiltonian including the pump field,
and $\mathcal{T}$ denotes time ordering. For the TrARPES spectra shown in the main text, the simulations use an in-plane pump electric-field of \(E_0 = 1.0 \times 10^8\) V/m at \(\hbar\omega = 310\) meV. Crystal symmetry near the probed K point constrains the photoemission matrix elements for the near-normal-incidence probe. With the crystalline $x$ axis along $\Gamma–K$, $x$- and $y$-polarized probes give $M_{A,\mathbf{k}}=M_{B,\mathbf{k}}$ and $M_{A,\mathbf{k}}=-M_{B,\mathbf{k}}$ at the K-point, respectively. 
Via the dynamical phase 
$
\mathcal V_{\mathbf p}(t,t')
=
\exp\left[
\frac{i}{\hbar}
\int_{t'}^t
d\bar t\,
\delta\epsilon_V(\mathbf p_\parallel,\bar t)
\right]
$,
final-state Volkov dressing\cite{stefan2025graphene} is included 
where 
$
\delta\epsilon_V(\mathbf p,t)
=
\frac{\hbar\vb p_\parallel}{m}
\cdot
e\mathbf A_V(t)
+
\frac{e^2A_V^2(t)}{2m}
$
is the pump-induced energy shift of the free space photo-electron.
Here $\vb p_\parallel=\vb k+\vb G$ is its in-plane momentum, $\vb k$ is the crystal momentum of the initial state and $\vb G$ is a reciprocal-lattice vector. The vector potential acting on the free space photoelectron is
$\vb A_V(t)=\alpha_V\vb A_{\rm pump}(t)$ where  $\alpha_V=0.7$ is a  phenomenological coefficient.
Note that  this effect  only dresses the free space photoelectron  and leads to Floquet--Volkov interference, while it does not affect the Floquet bands of the  Bloch  electrons.

\end{methods}

\section*{References}

\begin{addendum}
	\item[Data availability]
	All data supporting the results of this study are available within the article. Additional data are available from the corresponding author upon request.
	
  \item [Acknowledgement] 
  
  This work is supported by the National Natural Science Foundation of China (grant numbers 12421004 and 12234011), Tsinghua University Initiative Scientific Research Program (grant number 20251080106), National Key R\&D Program of China (grant number 2021YFA1400100), the National Natural Science Foundation of China (grant numbers 12327805, 52388201 and 12374291) and New Cornerstone Science Foundation through the XPLORER PRIZE. A.R. acknowledges support by the European Research Council (ERC-2024-SyG-101167294; UnMySt), the Cluster of Excellence Advanced Imaging of Matter (AIM), Grupos Consolidados Gobierno Vasco (IT1453-22), and the support from the Max Planck-New York City Center for Non-Equilibrium Quantum Phenomena. The Flatiron Institute is a division of the Simons Foundation. P.T. was supported by the National Key Basic Research and Development Program of China (Grants No. 2024YFA1409100), the National Natural Science Foundation of China (Grants No. 12234011 and No. 12374053), and the Suzhou National Laboratory (SK-1902-2026-004).
 
  \item[Author Contributions] 
  S.Z. conceived the research project. F.W. and X.C. performed TrARPES measurements and analyzed the data. W.C., Jinxi L. and Jiansong L. prepared the samples. X.C. contributed to the development and optimization of the HHG light source. Jinze L. and Z.S. performed the theoretical calculations. B.F., P.T., W.D., A.R. and P.Y. discussed the results. F.W., Jinze L., P.Y., Z.S., and S.Z. wrote the manuscript, and all authors commented on the manuscript.

  \item[Competing Interests] 
  The authors declare that they have no competing financial interests.

\end{addendum}
\newpage



\begin{figure*}[htbp]
\centering
\includegraphics[width=16.8 cm]{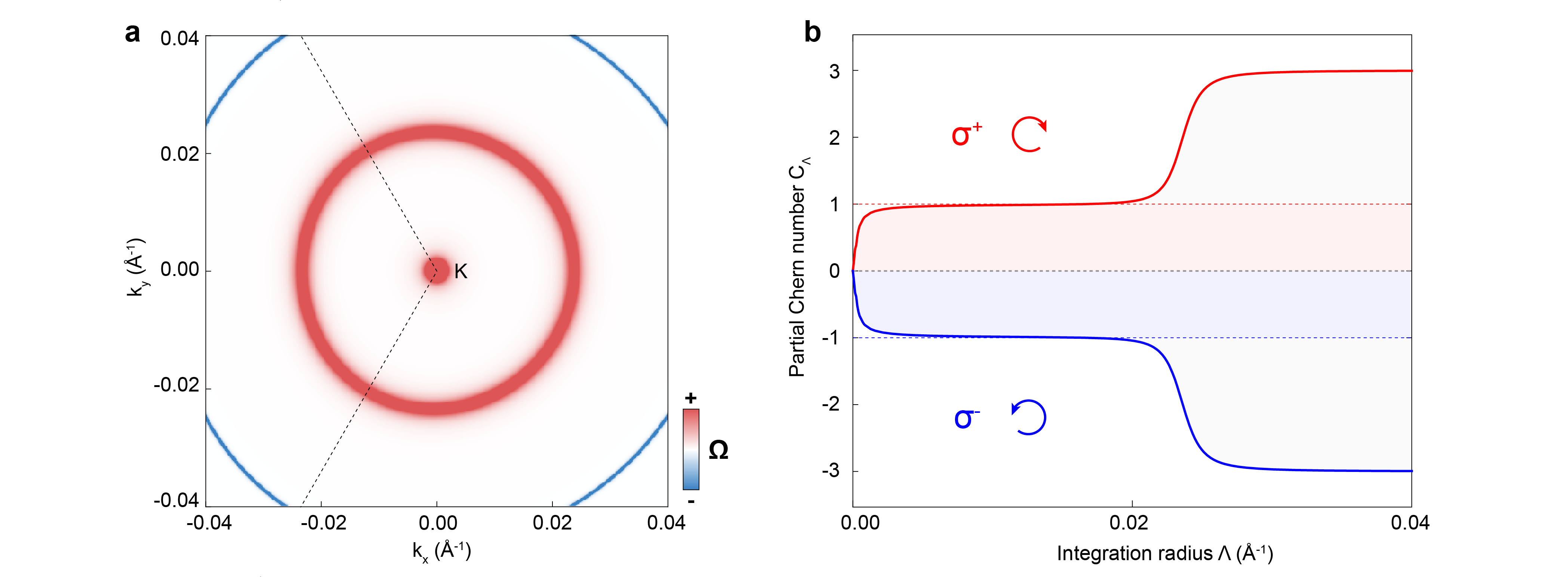}
\caption*{\textbf{Extended Data Fig.~1 ${\mid}$ Light-induced Berry curvature and Floquet Chern topology.} 
\textbf{a}, Calculated momentum-resolved Floquet Berry curvature of the lower Floquet band around the K valley.
\textbf{b}, The partial Chern numbers $C_\Lambda$ (including both valleys) as functions of the momentum cutoff $\Lambda$  for RCP (\(\sigma^+\)) and LCP (\(\sigma^-\)) with a weak pump field, respectively. 
}\label{FigS4}
\end{figure*}
\newpage

\newpage

\end{document}